\documentclass[12pt]{article}%
\usepackage{amssymb,amsmath,xcolor,graphicx,xspace,colortbl,ragged2e,rotating}
\usepackage{amsfonts}
\usepackage{amsmath}
\usepackage{amssymb}
\usepackage{boxedminipage}
\usepackage{graphics}
\usepackage{graphicx}
\usepackage{ragged2e}
\usepackage{wrapfig}
\usepackage{xcolor}%
\providecommand{\U}[1]{\protect\rule{.1in}{.1in}}
\graphicspath{{Fresnel-comment_graphics/}{Fresnel-comment_tcache/}{Fresnel-comment_gcache/}}
\DeclareGraphicsExtensions{.pdf,.eps,.ps,.png,.jpg,.jpeg}
\providecommand{\U}[1]{\protect\rule{.1in}{.1in}}
\begin{document}

\title{On Fresnel Aether Drag, `Moving' Images, and Relativity}
\author{C. S. Unnikrishnan\\\relax \textit{Tata Institute of Fundamental Research, } \\\relax \textit{Homi Bhabha Road, Mumbai - 400 005, India} \\\relax E-mail address: unni@tifr.res.in}
\date{}
\maketitle

\begin{abstract}
I show the decisive difference between genuine transverse Fresnel drag of
light in a moving medium and the \textquotedblleft spatial
shift\textquotedblright\ measured with a time dependent interference pattern
of light traversing a homogeneous finite medium (J. Leach \textit{et al.}, PRL
100, 153902 (2008)). In the latter case, the relative velocity and spatial
shift are in fact zero and the `movement' is an elementary visual illusion,
easily made superluminal. Three separate proofs are given for this fact. What
is recorded in the experiment is just the difference between a time dependent
space-fixed pattern and its time lagged version. This has no relevance to
relative motion of any physical entity, Fresnel drag or relativity.

\end{abstract}

Fresnel drag, first measured interferometrically by Fizeau, is an important
pre-relativity result on the propagation of light in moving media with phase
refractive index $n$ \cite{Fizeau}. When the velocity of the medium is
parallel to the propagation direction, the resultant velocity of light is
given by
\begin{equation}
c^{\prime}=\frac{c}{n}\pm v(1-1/n^{2})
\end{equation}
So, the drag velocity is $v_{d}=v(1-1/n^{2})\text{,}$ with the characteristic
and crucial $1/n^{2}$ dependence in the effective drag. The expression can be
derived from special relativity's velocity addition formula as well
\cite{Eins1905}.%
\begin{figure}[ptb]%
\centering
\includegraphics[
natheight=0.960300in,
natwidth=1.872700in,
height=1.0172in,
width=1.9545in
]%
{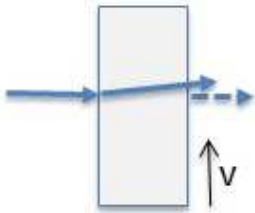}%
\caption{Transverse Fresnel Drag happens when the motion of the medium is in a
direction perpendicular to that of the light ray.}%
\label{TFdrag}%
\end{figure}

When light propagates in a direction transverse to the motion of the (finite)
medium, the `drag' should carry light laterally, resulting in a shift of point
of emergence of the ray (fig. 1). This transverse Fresnel drag was measured
carefully by the master metrologist R. V. Jones and confirmed that it follows
the relation
\begin{equation}
\delta x=\frac{vL}{c}(n_{g}-1/n_{p})
\end{equation}
where $L$ is the thickness of the medium, $n_{g}$ its group refractive index
and $n_{p}$ its phase refractive index \cite{Jones1,Jones2}. Since the time
for propagation inside the medium is $L/\left(  c/n_{g}\right)  \text{,}$
light is dragged by
\begin{equation}
\delta x=v_{d}t=\frac{vL}{c/n_{g}}(1-\frac{1}{n_{p^{2}}})\simeq\frac{vL}%
{c}(n_{g}-\frac{1}{n_{p}})
\end{equation}
This is a difficult measurement compared to the usual Fresnel drag because it
is not amenable to any straightforward interferometric scheme. The shift to be
measured, taking 10 m/s for the velocity of the medium, with total thickness 5
cm in a double pass configuration, is less than 2 nm. This needs to be
measured to the accuracy of 1\%, to compare with different theoretical
possibilities. This task was admirably achieved by Jones in two experiments,
requiring elaborate mechanical and optical arrangement \cite{Jones1,Jones2}%
.\qquad

More recently, Leach \textit{et al.} considered a special relativistically
symmetric situation of the light beam moving transversally in a lab-fixed
static medium \cite{Leach etal}. The desire was to measure the transverse drag
when there was \emph{relative motion} between the medium and light field, by
moving the light field across the medium that was static in the laboratory.
However, instead of moving the source of light and field relative to the
medium, they used the interference of two optical fields with slightly
different frequency, spatially overlapping at an angle. This results in a
pattern of fringes that are time dependent, visually mimicking a movement of
the pattern, though its spatial envelope is static. They reported a measured
drag and `spatial shift' of the pattern of light or the `optical image' that
has passed through the medium while transversely moving, following the
formula
\begin{equation}
\delta x^{\prime}=\frac{vL}{c}(n_{g}-1) \label{Leach}%
\end{equation}
instead of the Fresnel relation. In terms of the drag velocity, this is
equivalent to setting $v_{d}^{\prime}=v(1-1/n)\text{,}$ in contradiction to
the characteristic $1/n^{2}$ Fresnel drag. Quoting Leach \textit{et al.},
\textquotedblleft The discrepancy between our results and the work of Jones is
intriguing. For each configuration, either moving medium (Jones's experiment)
or moving image (our experiment), the analysis of the phenomenon is
explainable in either the rest frame of the medium or the frame in which the
medium is moving.\textquotedblright

The unexpected large deviation from the Fresnel drag relation was analyzed in
terms the difference between the Poynting vector along energy flows and the
wave vector, in an attempt to understand the discrepancy. However, a
supplementary experiment in which an optical interference pattern was rotated
inside the medium, also gave a result following eq. \ref{Leach}, instead of
the Fresnel drag relation that was expected. Such interference fringes are
formed by overlapping helically phased optical beams. This was considered
`puzzling', by Leach \textit{et al}.%
\begin{figure}[ptb]%
\centering
\includegraphics[
natheight=1.649500in,
natwidth=4.316100in,
height=1.2324in,
width=3.1913in
]%
{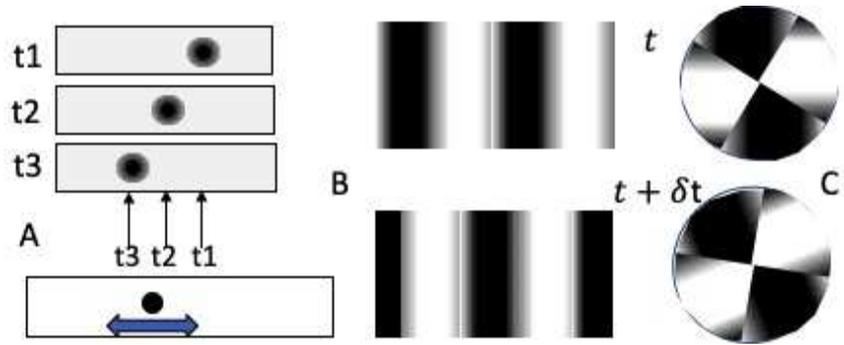}%
\caption{The visual illusion of transverse movement is created by a time
depended image field. A) A dark stop moves across a bright field, creating a
`moving' shadow, but the photons (or their absence) are not in relative
motion. B) Time dependent phase difference of two light beams with different
frequencies creates the illusion of the `moving fringes'. The photons do not
have a transverse velocity. C) Similar situation with cylindrical symmetry,
where are phase gradient is helical.}%
\label{illusion}%
\end{figure}

I now show that what was measured in the experiment by Leach \textit{et al.}
has no relevance for relative motion, Fresnel drag and relativity. All the
results of Leach \textit{et al.} are explained as the comparison of a time
dependent intensity pattern with its time delayed copy, without any genuine
spatial movement or velocity. A concise version of the essential argument was
published as a short comment \cite{Unni-prl-com}. Moving an aperture across a
light field results in the visual impression of a bright spot of light falling
on a transverse surface as `moving' (fig. 2A). The illusion of motion is
generated by `moving' images in the static visual frame, as in cinema; there
is no transverse motion of light. The method used by Leach \textit{et al.} is
effectively the same kind of visual illusion -- animation rather than motion.

Relativity deals with the relations between physical entities in two frames
that are relatively moving. One common situation is when the medium is moving
relative to the source of the optical field. To see what the reciprocal
situation is, one has to just shift to the frame of the moving medium.
Relative to that frame, the source of the optical field is moving (even the
lab and rest of the world is, but we need not consider that in the present
analysis). Then the photons are expected to have a transverse velocity,
relative to the medium. This is not what was done in the experiment. Leach
\textit{et al.} created the `impression' of lateral movement of an optical
field relative to a medium -- a block of glass -- by introducing a frequency
difference between two beams overlapping at a small angle resulting in a time
dependent phase at a transverse field point $(x,y)\text{,}$ but no relative
motion at any velocity was involved. When wavefronts overlap at small angle
$2\alpha$, parallel fringes with spacing $\Lambda=\lambda/2\alpha$ are formed
and a frequency difference $\delta\nu$ between the beams create the impression
of moving fringes with apparent velocity $v_{a}=\Lambda\delta\nu\text{,}%
$without the beams moving (fig. 2B). That it is not a physical motion in the
sense of relativity and kinematics is easily seen by considering beams of size
5 cm, with a small angle between them $\alpha\simeq2\times10^{-5}$rad,
$\Lambda\simeq2.5$ cm, and a practical $\delta\nu\simeq20$ GHz. The `velocity'
then is an unphysical $5\times10^{8}$ m/s! (one can also just magnify the
fringe pattern and then the `velocity' increases unphysically with magnification).

The fact that the optical impression is not physically relevant motion can be
illustrated easily, and in many ways. If the interference fringes are formed
such that the visibility is not 100\%, then the whole pattern will not change
across the field of view. The progressive time dependent phase change `moves'
only the partial fringe pattern. The rest of the light forms an incoherent
static field. Hence, it is obvious that the `movement' is mere optical illusion.

Another \textit{experimental demonstration} is the following. If the visual
impression were genuine motion, the photons would have a transverse velocity
relative to lab-fixed references. With fringe spacing of 1 cm and a frequency
difference of 30 MHz, the transverse velocity is $3\times10^{5}\text{{}}$ m/s.
If the light in the pattern is passed through a small aperture, this stream of
photons would reach a screen at a distance $L=1\text{{}}$ m with a transverse
shift equal to $\Delta x=Lv/c\simeq1$ mm. This is easily measurable with a CCD
camera. However, it may be verified that there is no such shift. Thus, I have
mentioned three different proofs for the illusion of relative motion in the
experiments by Leach \textit{et al.}

All these comments apply to the motional animation involving circularly
symmetric optical patters as well, used in the experiments by Leach \textit{et
al}. There, the linear patterns looped in a circle generate the visual
impression of circular movement.

The relative phase of the two beams at nearby points in the transverse plane
change by
\begin{equation}
\frac{d\left(  \delta\phi(x)\right)  }{dt}=\delta\nu\frac{4\pi\alpha\delta
x}{\lambda}%
\end{equation}
but clearly there is no transverse motion of any physical entity. The optical
field at $x^{\prime}$ is a copy of the field at $x$ at an earlier time $\delta
t=2\alpha\left(  x^{\prime}-x\right)  /\lambda\delta\nu$ without any
transverse component of velocity imparted to the optical beam. There is no
velocity or photon momentum in the transverse direction. In fact, one may
consider just one of the interfering beams. If we move a periodic grid in
front of the beam, the shadow (the contract between the bright and dark
regions) moves across at an apparent velocity \emph{but there is no movement
of the beam itself nor any relative velocity} \emph{between the optical beam
and the medium} in the sense of relativity and kinematics. This will give the
same results obtained by Leach\textit{ et al}. The case of interference
pattern is similar, with one more optical field added to the configuration,
since the light beams are not moving relative to the medium or the lab. Now,
if we compare this time dependent intensity with itself after a time delay
$\delta t\text{,}$ without any spatial drag or displacement, we get the
difference
\begin{equation}
\delta I(x)=\delta t\times\frac{\partial I(x)}{\partial t}%
\end{equation}
The time delay in this case is due to the two different paths to the CCD
camera, one through free space and another through glass of thickness
$L\text{.}$
\begin{align}
\delta t  &  =\frac{L}{c/n}-\frac{L}{c}=\frac{L}{c}(n-1)\\
\delta I(x)  &  =\frac{\partial I(x)}{\partial t}\frac{L}{c}(n-1)
\end{align}
So, the total beam at two different times occupies exactly the same spatial
region and boundary, but the internal intensity pattern across the beam $I(x)$
differ by the temporal `speed' at which the intensity is modulated, without
genuine motion or velocity. This is what Leach \textit{et al.} saw in their
experiments, both in the linear version and in the rotational version. Because
the experiment has no relation to movement in relativity or in Fresnel drag,
the experiment does not address those relevant issues. The camera is comparing
the time dependent spatial intensity pattern with an earlier copy, arrived
delayed through glass, giving the false impression of a spatial shift.
\emph{That there is zero spatial shift can easily be seen from the boundary of
the image or the beam, which is part of the optical field and remains static
relative to the medium.} This lag could have been just an optical delay in
free space without a medium, resulting in two paths with a delay between them
and the same result would have followed, which is
\begin{equation}
\delta x^{\prime}=v\frac{L_{1}-L_{2}}{c}=v\delta t
\end{equation}

The experiment with rotating image is identical. An time dependent, \emph{but
spatially stationary} image is compared with its slightly earlier copy and the
difference in angles will of course be
\begin{equation}
\delta\theta^{ \prime} =\frac{\Omega L}{c} (n_{g} -1)
\end{equation}
instead of the Fresnel result
\begin{equation}
\delta\theta=\frac{\Omega L}{c} (n_{g} -\frac{1}{n_{p}})
\end{equation}

This completely explains the \emph{non-relativity results} in the paper by
Leach \textit{et al}., obtained due to the use of motion-animation rather than
genuine motion, .

In fact, the apparent movement in space is not important. The modulation can
be in any quantity associated with beam, like the frequency, polarization
etc., and two measurement with a time delay generated with a medium will
differ by the same formula, with the $L (n -1)/c$ dependence. If the image was
changing color linearly (frequency of light) at some rate, the difference
between the two images will be in frequency space with a shift
\begin{equation}
\delta f^{ \prime} =\frac{\left(  d f/d t\right)  L}{c} (n_{g} -1)
\end{equation}
If polarization was changing linearly in time, then `shift' in polarization,
\[
\delta\varepsilon^{ \prime} =\frac{\left(  d \varepsilon/d t\right)  L}{c}
(n_{g} -1)
\]

The fine measurements done by R. V. Jones on the transverse Fresnel drag in
1970s are yet to be surpassed in precision and ingenuity. Measurements using
medium with very large refractive index is not very useful because large group
refractive index $n_{g}$ renders the crucial `relativistic' term $-1/n_{p}%
^{2}$ unobservable. Measurements in moving media with anomalously large
effective refractive index, like what can be arranged with an EIT
(Electromagnetically Induced Transparency) medium, are not relevant to the
issues of relativity! This is because, the Fresnel drag term $(1-1/n^{2})$ is
totally negligible compared to the anomalous refractive index, $\Delta
n=\left(  \partial n/\partial\omega\right)  \omega/n$. What Fizeau and Jones
achieved was not just the measurement of the drag of light by the medium, but
the verification of the crucial fact of the `partial drag', which is the
relativistic signature. Complete drag is Galilean with no indication of a
limiting velocity. Therefore, the measurement of a genuine relativistic
Fresnel drag is a problem in the domain of precision measurements requiring
considerable ingenuity. The specific issue of whether the drag is symmetric
between the movement of the medium and movement of the light field is a more
difficult issue to answer experimentally. Possibilities for new measurements
will be discussed in another paper, in a wider context covering optics,
quantum mechanics and relativity \cite{Unni-Fresnel2019}.

\end{document}